\documentclass{appolb}
\usepackage{epsfig}

\def\kk{ {KK} }
\def\beq{\begin{equation}}
\def\eeq{\end{equation}}
\def\bea{\begin{eqnarray}}
\def\eea{\end{eqnarray}}
\def\nn{\nonumber}

\def\roughly#1{\mathrel{\raise.3ex\hbox
{$#1$\kern-.75em\lower1ex\hbox{$\sim$}}}}
\def\lsim{\roughly<}

\def\sss{\scriptscriptstyle}

\def\bs{B_s^0}

\def\kbar{{\bar K}^0}

\def\ANPu{{\cal A}^u}
\def\ANPd{{\cal A}^d}

\def\bskk{{\bs}\to K^+ K^-}
\def\bskkneut{{\bs}\to K^0 \kbar}

\begin{document}
\title{ Merging Flavour Symmetries with QCD Factorisation for $B\to KK$ Decays.%
\thanks{Presented at Final Euridice Meeting, Kazimierz, August 2006}%
}
\author{Joaquim Matias
\address{IFAE, Universitat Aut\`onoma de Barcelona, 08193 Bellaterra, Barcelona, Spain}
} \maketitle
\begin{abstract}
The interplay between flavour symmetries connecting $B_s \to KK$
decays with the recently measured $B_d \to K^0 {\bar K}^0$ decay and QCD
Factorisation opens new strategies to describe the decays $B_s \to
K^0 {\bar K}^0$ and $B_s \to K^+ K^-$ in the SM and in
supersymmetry. A new relation, emerging from the sum-rule for the
$B_s \to K^0 {\bar K}^0$ decay mode, is presented  offering a new way to
determine the weak mixing angle $\phi_s$ of the $B_s$ system.
\end{abstract}

\section{Introduction}
The huge effort on the experimental side at present B facilities
(Babar, Belle and CDF) to increase the precision on data measurements
force us to revise the strategies on the theory side to produce
more accurate predictions. Non leptonic B decays offer different
strategies to determine the Unitarity Triangle, to search
for New Physics (NP)\cite{np} but also to rule out models \cite{fbm}. While a lot of attention has been devoted to
the $B\to \pi K$\cite{bpk1,bpk2,bpk3,bpk4,FLBUR} decay modes, here we will focus on $B\to KK$ decays that has
been observed at CDF \cite{cdf} ($B_s \to K^+ K^-$) and at Babar
\cite{babar} and Belle \cite{belle} ($B_d \to K^0 {\bar K}^0$).

There are two main approaches in the literature to describe
$B \to KK$ decays: flavour symmetries and $1/m_b-$expansion methods (QCD
Factorisation~\cite{BBNS,BN}/ soft collinear effective
theories~\cite{scet} or PQCD \cite{pqcd}). Each of those methods has pros and
cons, that we will discuss in turn. Flavour symmetries, like
U-spin symmetry  that relates $B_s \to K^+K^-$ with $B_d
\to \pi^+ \pi^-$~\cite{fl,LM,LMV,LMV2}, provide a model independent analysis and
extract most of the needed hadronic parameters from data. However,
this method has the disadvantage that it relies strongly on the
accuracy of data, and, at present, there is still not full agreement between
Babar and Belle data on the CP
asymmetries of the $B_d \to \pi^+ \pi^-$ mode. As a consequence, error bars are still
quite large, see for instance, the prediction  ${\rm BR}(B_s
\to K^+K^-)=\left(35^{+73}_{-20}\right) \times 10^{-6}$ \cite{FLBUR} or, more recently,  $4.2 \times 10^{-6} \leq {\rm BR}(B_s
\to K^+K^-) \leq 60.7 \times 10^{-6}$~\cite{LMV2}. Also
when relating $B_d \to \pi^+ \pi^-$  with
 $B_s \to K^+K^-$  some of the needed U-spin
parameters can only be roughly estimated and they are usually taken to be of the order of $20 \%$.

Concerning $1/m_b$-expansion methods, here we will focus on QCD
Factorisation (QCDF)~\cite{BBNS,BN,B}. The main idea  is to exploit the existence of a
large scale $m_b \gg \Lambda_{QCD}$ together with colour
transparency, that applies when the outgoing meson that does not
contain the spectator quark is very energetic. At leading power in
$\Lambda/m_b$ all long distance contributions can be parametrized
in terms of form factors and light cone distribution amplitudes, while the
contribution from energetic gluons comes in a perturbative series
in $\alpha_s$ and it is incorporated into the hard scattering
kernels. QCDF predicts some of the hadronic
parameters reducing the error bars, however in the computation one
has to face chirally enhanced IR divergences.
They are formally suppressed by a power of $1/m_b$,
   but can be numerically significant. They are modelled
and induce an important uncertainty in the predictions.

However, there is a third possibility and it is the proposal
presented in \cite{dmv} that combines QCDF and Flavour symmetries
giving rise to rather accurate predictions for the branching
ratios of the above mentioned $B_s \to KK$ decays in SM and in
supersymmetry. Moreover, the method predicts some of the SU(3)
breaking parameters which can be useful for other flavour
approaches and, at the same time, deals with the problem of the
chirally enhanced  IR divergences coming at order $\Lambda/m_b$ (see also \cite{fh}).

Since IR divergences play a central role in this discussion, it is
worth to mention  the two sources of IR divergences in QCDF:

\begin{itemize} \item Hard spectator-scattering: Hard gluons exchange
between spectator quark and the outgoing energetic meson gives
rise to integrals of the following type  (see \cite{BN} for definitions):
$$
   H_i(M_1M_2)
   = C
   \int_0^1\!dx \int_0^1\!dy \left[
   \frac{\Phi_{M_2}(x)\Phi_{M_1}(y)}{\bar x\bar y}
   + r_\chi^{M_1}\,\frac{\Phi_{M_2}(x)\Phi_{m_1}(y)}{x\bar y} \right], \nonumber
$$
where the second term (formally of order $\Lambda/m_b$) diverges
when $y\to 1$.

\item Weak annihilation: These type of diagrams also
exhibit endpoint IR divergences as it is explicit in the corresponding integrals: \begin{eqnarray} A_1^i =
\pi\alpha_s \!\!\!\!\!\!\!\!\!&&\int_0^1 dx dy\,
    \left\{  \Phi_{M_2}(x)\,\Phi_{M_1}(y)
    \left[ \frac{1}{y(1-x\bar y)} \right. \right.
     \nonumber \cr &+& \left. \left.
    \frac{1}{\bar x^2 y} \right]
    + r_\chi^{M_1} r_\chi^{M_2}\,\Phi_{m_2}(x)\,\Phi_{m_1}(y)\,
     \frac{2}{\bar x y} \right\}. \end{eqnarray}

\end{itemize}
Both divergences are modelled in the same way \cite{BN}:
$$
   \int_0^1\frac{d y}{\bar y}\Phi_{m_1}(y)
   \!=\! \Phi_{m_1}(1)\int_0^1\frac{d y}{\bar y}
    + \int_0^1\frac{d y}{\bar y}\Big[ \Phi_{m_1}(y)-\Phi_{m_1}(1)
    \Big]
  \!\equiv\! \Phi_{m_1}(1)\,X_{H,A}^{M_1}
    +  {r,} \nonumber $$
where $r$ is a finite piece and the divergent piece is cut-off by a physical scale of order
$\Lambda_{QCD}$ with an arbitrary complex coefficient to take into
account possible multiple soft scattering: $X_{H,A}=(1+
\rho_{H,A})\, {\rm ln} (m_b/\Lambda)$.

\section{Sum-rules: $\alpha$ and $\phi_s$ }

The SM amplitude for a $B$-decay into two mesons can be split into
tree and penguin contributions~\cite{B}\footnote{Conventionally,
we will call ``tree" the piece proportional to $\lambda_u^{(q)}$
and ``penguin" the piece proportional to $\lambda_c^{(q)}$, even
if applied to decays with no actual tree diagram.} :$
\bar{A}\equiv A(\bar{B}_q\to M \bar{M})
  =\lambda_u^{(q)} T_M^{qC} + \lambda_c^{(q)} P_M^{qC}\,,
$ with $C$ denoting the charge of the decay products, and the
products of CKM factors $\lambda_p^{(q)}=V_{pb}V^*_{pq}$.
Tree and penguin contributions in $\bar{B}_d\to K^0 \bar K^0$ in
QCDF are:
\begin{eqnarray}
{\hat T^{d\, 0}} &=&
  \alpha_4^u-\frac{1}{2}\alpha_{4EW}^u
    +\beta_3^u +2 \beta_4^u - \frac{1}{2} \beta^u_{3EW} - \beta^u_{4EW}
    \nonumber
\\  \label{eq4}
{\hat P^{d\, 0}} &=&
 \alpha_4^c-\frac{1}{2}\alpha_{4EW}^c
    +\beta_3^c +2 \beta_4^c
    - \frac{1}{2} \beta^c_{3EW} - \beta^c_{4EW},
\end{eqnarray}
where $\hat P^{d0}=P^{d0}/A^d_\kk$, $\hat T^{d0}=T^{d0}/A^d_\kk$,
the super-scripts identify the channel,  the normalisation
 is $A^q_\kk=M^2_{B_q} F_0^{\bar{B}_q\to K}(0) f_K
{G_F}/{\sqrt{2}}$ (see \cite{BN,dmv} for the corresponding
expression of the $B_s \to KK$ channels). Following the
observation in \cite{dmv} that the structure of the IR divergences
is the same, independently of the charm or up quark running in the
loop, we identified an IR-safe quantity at NLO in QCDF  that we
called $\Delta_d\equiv T^{d0}-P^{d0}$. All
chirally enhanced IR divergences cancel exactly in this quantity at this order. Its
explicit expression in terms of the coefficients in Eq.(\ref{eq4}) is:
$$
\Delta_d=A_{kk}^d[\alpha_4^{u}-\alpha_4^{c} +
\beta_3^{u}-\beta_3^{c}+2\beta_4^{u}-2\beta_4^{c}],
$$
where electroweak contributions are neglected. This quantity can be safely evaluated in QCDF and the result found
in \cite{dmv} was: $ \Delta_d=(1.09\pm 0.43) \cdot 10^{-7} + i
(-3.02 \pm 0.97) \cdot 10^{-7} {\rm GeV}$. The largest uncertainty
entering $\Delta_d$ comes from the ratio $m_c/m_b$ and the scale
dependence. Interestingly, this quantity can be expressed in terms
of observables, providing a relation between the direct induced
CP-asymmetry ($A_{dir}^{d0}$), the mixing induced CP-asymmetry
($A_{mix}^{d0}$) and the branching ratio ($BR^{d0}$) of
$\bar{B}_d\to K^0 \bar K^0$:
$$ \label{eq:srd}
|\Delta_d|^2\!=\!{\frac{BR^{d0}}{L_d}} \{x_1 + [x_2 \sin\phi_d -
x_3 \cos\phi_d] A_{mix}^{d0}
   -[x_2 \cos\phi_d + x_3 \sin\phi_d ] A_{\Delta}^{d0} \}\,, \nonumber
$$
where $|A_\Delta^{d0}|^2+|A_{\rm dir}^{d0}|^2+|A_{\rm
mix}^{d0}|^2=1$, $\phi_d$ is the weak mixing angle for the $B_d$
system, $L_d=\tau_d \sqrt{M_{Bd}^2 - 4 M_K^2}/(32 \pi M_{Bd}^2)$
 and $x_i$ are functions of $\gamma$ and CKM elements. All SM inputs are taken as in
\cite{dmv} following \cite{ckmfitter}.

Moreover, it was found in \cite{dlim} that this sum-rule encodes
also a very interesting information. It provides a new way of
measuring $\sin \alpha$:
$$ \label{eq:srd}
 \sin^2 \alpha= \frac{BR^{d0}}{4 L_d |\lambda_u^{(d)}|^2
|\Delta_d|^2}
\left(1-\sqrt{1-{A_{dir}^{d0}}^2-{A_{mix}^{d0}}^2}\right).
 \nonumber
$$
(See \cite{zupan} for a recent review on the extraction of $\alpha$).
In a similar way, it was found in \cite{dmv} a corresponding
IR safe quantity $\Delta_s\equiv T^{s0}-P^{s0}$ and sum-rule for the
decay $\bar{B}_s\to K^0 \bar K^0$:
$$ \label{eq:srd} |\Delta_s|^2\!=\!{\frac{BR^{s0}}{L_s}} \{y_1
+ [y_2 \sin\phi_s - y_3 \cos\phi_s] A_{mix}^{s0}
   -[y_2 \cos\phi_s + y_3 \sin\phi_s ] A_{\Delta}^{s0} \}\,, \nonumber
$$
which provides a completely new way to determine the weak mixing
angle $\phi_s$ that we present here (see \cite{dmv} for definitions):

$$ \label{eq:srd} \sin^2 \frac{\phi_s}{2}= \frac{BR^{s0}}{4 L_s
|\lambda_c^{(s)}|^2 |\Delta_s|^2}
\left(1-\sqrt{1-{A_{dir}^{s0}}^2-{A_{mix}^{s0}}^2}\right).
 \nonumber
$$
This implies that a measurement of the branching ratio, direct CP
asymmetry and mixing induced CP asymmetry of the decay $B_s
\to K^0 {\bar K}^0$ automatically translates into a value for
$\sin^2 \phi_s/2$. Finally, given the relation $\Delta_s=f
\Delta_d$, where $f=A^{s}_{KK}/A^{d}_{KK}$ a new relation between $\sin \alpha$ and
$\sin \phi_s/2$ immediately emerges.

\section{Description of the Method: Flavour Symmetries \& QCDF}
\subsection{$B_s \to K^0
{\bar K}^0$}

 The SM amplitude of this $b \to s$ penguin decay is
given by:
$${\bar A}\equiv A(\bar{B}_s\to K^0 \bar{K^0})
  =\lambda_u^{(s)} T^{s0} + \lambda_c^{(s)} P^{s0}.\, \nonumber
$$ Its dynamics is described in terms of three parameters: $|T^{s0}|$, $|P^{s0}|$
and the relative strong phase $\arg(P^{s0}/T^{s0})$ (remember that
$T^{s0}$ stands for the piece proportional to $\lambda_u^{(s)}$
but it is not due to an actual tree diagram in this case). Its hadronic
parameters can be related via U-spin with the hadronic parameters
($|T^{d0}|$, $|P^{d0}|$ and $\arg(P^{d0}/T^{d0})$) of the also penguin
governed mode $B_d\to K^0 \bar{K^0}$. This has several
advantages: first, we can expect similar final state interactions
(although not equal), second, the sources of U-spin breaking can be better
controlled using QCDF. These sources are: i) the factorisable ratio
$f={A_{KK}^s}/{A_{KK}^d}
\nonumber $ (extrapolated from the lattice) ii) U-spin breaking
$1/m_b$ suppressed terms $\delta \alpha_i$ and $\delta \beta_i$: sensitive to the
difference of $B_d$ and $B_s$ distribution
amplitudes  and spectator quark dependent
contributions coming from a gluon emitted from the $d$ or $s$
quark. This leads to the relations $|P^{s0}/(fP^{d0})-1| \leq 3\%$
and $|T^{s0}/(fT^{d0})-1| \leq 3 \% $.

The next step is to determine the hadronic parameters ($T^{d0}$,
$P^{d0}$) of the decay $\bar{B}_d\to K^0 \bar{K^0}$. This is done
using as inputs the ${\rm BR}(\bar{B}_d\to K^0 \bar{K^0})$ and
$\Delta_d$ (from QCDF). The direct CP asymmetry
$A_{dir}(\bar{B}_d\to K^0 \bar{K^0})$(denoted by $A_{dir}^{d0}$)
will be taken as a free parameter. The combination of those
constraints gives rise to a set of non-linear equations (see
definitions in \cite{dmv}):
\begin{eqnarray} \label{circ}
x_C+i y_C &=& -\Delta_d (1-{\cos\gamma}/{R})/a \,,\nonumber \\
r^2&=&{\rho_0^2}/[{a|\lambda_u^{(d)}|^2}]-[{\sin\gamma
|\Delta_d|}/({aR})]^2 \,, \nonumber \\  y_P
x_{\Delta}&=&y_{\Delta}x_P - {\rho_0^2
A_{dir}^{d0}}/({2|\lambda_u^{(d)}\lambda_c^{(d)}|\sin\gamma})\,,
 \end{eqnarray}
 that determines $P^{d0}=x_P+i y_P$, then using $\Delta_d$
one gets $T^{d0}$. Two remarks are important here: first, there is a
twofold ambiguity in the sign of ${\rm Im} P^{d0}$ (solved in the next
subsection). Second, current data together with
our knowledge on $\Delta_d$  limits the
 $A_{dir}^{d0}$  asymmetry (by means of Eqs.(\ref{circ})) within a restricted range
between $-0.2 \leq A_{dir}^{d0} \leq 0.2 $.

Finally, our SM predictions for the branching ratio and CP
asymmetries of $\bar{B}_s\to K^0 \bar{K^0}$ are obtained using the
relations between $P^{s0} \leftrightarrow P^{d0}$ and $T^{s0} \leftrightarrow T^{d0}$
mentioned above and including all U-spin breaking sources together
with the QCDF uncertainties in $\Delta_d$. The resulting
predictions are $BR(B_s \to K^0 \bar{K^0})=(18 \pm 7
\pm 4 \pm (2)) \times 10^{-6}$, $|A_{dir}(B_s \to K^0 {\bar K^0})|\leq1.1 \times 10^{-2}$ and
$|A_{mix}(B_s \to K^0 {\bar K^0})|\leq1.5 \times 10^{-2}$. The sign of these asymmetries
can be fixed once  $A_{dir}^{d0}$ will be measured with enough accuracy.

\subsection{$B_s \to K^+{K}^-$}

The analysis of the decay mode $B_s \to K^+K^-$ follows similar steps,
but with some important differences: i) we will
use U-spin and isospin to connect $B_d \to K^0 {\bar K}^0$ with
$B_s \to K^+K^-$, ii) $B_s \to K^+ K^-$ contains a tree (denoted by
$\bar \alpha_1$ in QCDF) with no counterpart in $B_d \to K^0 {\bar
K}^0$, iii) $BR(B_s \to K^+ K^-)$ has been measured with excellent
precision at CDF\cite{cdf} and iv) we will use the information only
on the {\it sign predictions } (not the absolute value) for the CP-asymmetries of $B_s \to K^+ K^-$ from the
strategy that uses U-spin to relate $B_s \to K^+ K^-$ with $B_d
\to \pi^+ \pi^-$.

The hadronic parameters describing the amplitude:
$$A(\bar{B}_s\to K^+ {K^-})
  =\lambda_u^{(s)} T^{s\pm} + \lambda_c^{(s)} P^{s\pm}\, \nonumber
$$
are obtained from the relations containing the sub-leading $1/m_b$
U-spin breaking:
$|P^{s\pm}/(fP^{d0})-1| \leq 2 \%$,
$|T^{s\pm}/(A^s_{kk} \bar\alpha_1)-1-T^{d0}/(A^d_{kk} \bar\alpha_1)| \leq  4 \%$.
Those errors, estimated within QCDF, are stretched roughly by a factor two to be
conservative.

The two-fold ambiguity on the sign of ${\rm Im}P^{d0}$ is lifted
here using U-spin arguments based on the $B_d \to \pi^+\pi^-$
strategy. While the signs of ${\rm Im} P^{d0}$ and $A_{dir}^{s\pm}$ are correlated
(being both positive or negative),
 the
prediction based on the $B_d \to \pi^+\pi^-$ strategy points
towards a positive sign for $A_{dir}^{s\pm}$\cite{LMV2}, discarding then the
solution with ${\rm Im}P^{d0}<0$.

Our result in the SM  for the branching ratio of $B_s \to K^+ K^-$
averaging over all values  of
$A_{dir}^{d0}$ is $BR(B_s \to K^+ K^-)=(20 \pm 8 \pm 4 \pm
(2))\times 10^{-6}$, where the last error in parenthesis stands
for a rough estimate of finite, non-enhanced $\Lambda/m_b$
corrections.
Finally, confronting our predictions for $B_s \to K^+ K^-$ with
the data on $B_d \to \pi^+ \pi^-$~\cite{data}: $\vert
T^{d\pm}_{\pi\pi}\vert=(5.48\pm 0.42)\times 10^{-6}$ and
$\left\vert {P^{d\pm}_{\pi\pi}}/{T^{d\pm}_{\pi\pi}} \right\vert
=0.13\pm 0.05$, $
\arg{\left({P^{d\pm}_{\pi\pi}}/{T^{d\pm}_{\pi\pi}}
\right)}=(131\pm 18)^\circ$, provides a double information. First,
we can give predictions for the
U-spin breaking parameters:
$\mathcal{R}_\mathcal{C}=|T^{s\pm}/T^{d\pm}_{\pi\pi}|=2.0\pm 0.6
$ and
$\xi=|P^{s\pm}/T^{s\pm}|/|P^{d\pm}_{\pi\pi}/T^{d\pm}_{\pi\pi}|=0.8\pm
0.3$ connecting $B_s \to K^+ K^-$ with $B_d \to \pi^+ \pi^-$. These parameters can be compared
 with the
QCD sum rules predictions in ref. \cite{khod}. Notice that while QCD sum
rules gives only
the factorizable part, our predictions include, in principle, the full contribution.
Second, a comparison between the two relative strong phases
$\arg\left(P^{s\pm}/T^{s\pm}\right)$ and
$\arg{\left({P^{d\pm}_{\pi\pi}}/{T^{d\pm}_{\pi\pi}}\right)}$ selects
$A_{dir}^{d0}\geq 0$. Then, if we restricts only to positive values of
$A_{dir}^{d0}$ according to the previous arguments, our SM
predictions turn out to be \cite{dmv,Baek}:
\begin{eqnarray} BR(B_s \to K^+
K^-)&\!\!\!\!=\!\!\!\!&(17 \pm 6 \pm 3 \pm (2))\times 10^{-6}, \nonumber \\
-0.22 \leq A_{dir}^{s\pm} \leq 0.49  &{\rm and}& -0.55 \leq
A_{mix}^{s\pm} \leq 0.02.
\end{eqnarray}
Another argument in favour of
$A_{dir}^{d0}\geq 0$ comes from the preference of $A_{mix}^{s\pm}<0$ of
 the U-spin based $B_d \to \pi^+ \pi^-$ strategy \cite{LMV} (see the anti-correlation between
 $A_{dir}^{d0}$ and $A_{mix}^{s\pm}$ in Table 1 of \cite{dmv}).

The accuracy on these CP-asymmetries will be substantially improved
once a precise measurement on $A_{dir}^{d0}$ will be available or
the error of  $BR(B_d \to K^0 {\bar K}^0)$ and the QCD
uncertainties on $\Delta_d$  (mainly $m_c/m_b$ and scale
dependence) will be reduced.

\subsection{Supersymmetry}

The leading gluino-squark box and penguin contributions \cite{bpk1} to $B_s
\to K^0 {\bar K}^0$ and $B_s \to K^+ K^-$ were evaluated first in
\cite{LMV2} using the U-spin flavour strategy with $B_d \to \pi^+
\pi^-$ and, afterwards, in \cite{Baek} using the new method
combining flavour and QCDF \cite{dmv}. The relative size of this
contribution compared to the SM penguin is $
({\alpha_s/M_{susy}^2})/({\alpha/M_W^2}) \sim 1$. The amplitude
of these decays in presence of NP contains an extra
contribution:
$ {\cal A}(\bskk)={\cal A}^{s\pm}_{\sss SM} +
\ANPu e^{i \Phi_u}$, ${\cal A}(\bskkneut) = {\cal
A}^{s0}_{\sss SM} + \ANPd e^{i \Phi_d}$.
These NP amplitudes $\ANPu e^{\Phi_u}$ and $\ANPd
e^{\Phi_d}$ in terms of  Wilson coefficients are:
\begin{eqnarray}
\mathcal{A}^q e^{i\Phi_q}\!=\!\frac{G_F}{\sqrt{2}} \Big[
-\chi(\frac{1}{3}\bar{c}_1^q+\bar{c}_2^q)-\frac{1}{3}(\bar{c}_3^q-\bar{c}_6^q)-(\bar{c}_4^q-\bar{c}_5^q)
-\lambda_t \frac{2\alpha_s}{3\pi}\bar{C}_{8g}^{\rm eff}\big( 1+\frac{\chi}{3} \big) \Big]A\nn
\end{eqnarray}
with $q=u,d$, $A=i (m_B^2-m_K^2) f_K F^{B_s \to K}$ and $\chi=1.18$ (see \cite{Baek} for definitions).
These Wilson coefficients are sensitive to the ${\tilde s}-{\tilde
b}$ mass splitting. After including the constraints coming from
$BR(B\to X_s \gamma)$, $B\to \pi K$ and $\Delta M_s$ we found that
a large isospin violation controlled by the mass splitting
$\widetilde{u}_R$-$\widetilde{d}_R$ is possible between the NP
amplitudes $\ANPu e^{i \Phi_u}$ and $\ANPd e^{i \Phi_d}$. For the
region of parameters considered in this supersymmetric scenario,  $\ANPu e^{i \Phi_u}$
can be up to
a factor three larger than $\ANPd e^{i \Phi_d}$. The specific results in supersymmetry for
each decay mode are \cite{Baek}:

\begin{itemize}

\item $\bskk$: The branching ratio  is very little
affected by SUSY. At most, the SM prediction can be increased by
15\% for $A_{dir}^{d0}=0.1$, increasing a bit the already good
agreement with the new CDF data~\cite{cdf}. The direct CP
asymmetry within SUSY falls inside the range $-0.1 \lsim
A_{dir}(\bskk)^{SUSY} \lsim 0.7$ for $-0.1 \leq A_{dir}^{d0} \leq
0.1$. The deviation depends on the relative size of the competing SM tree
versus the NP amplitude. $A_{mix}(\bskk)^{SUSY}$ can
take any value from [-1,1].

 \item $\bskkneut$: The impact of SUSY on $BR(\bskkneut)$ is even smaller,
reflecting the reduced allowed region for $\ANPd e^{i\Phi_d}$ as
compared to $\ANPu e^{i \Phi_u}$. The situation is very different
for the CP asymmetries, that are particularly promising, due to
the very small size  of their SM prediction
\cite{dmv}. The direct CP asymmetry in SUSY can be 10 times larger
than the SM one.
$A_{mix}(\bskkneut)^{SUSY}$ covers the entire range, and so this
asymmetry can be large in the presence of SUSY, contrary to the SM
prediction.

\end{itemize}

\noindent The method discussed here
is being applied to other non-leptonic $B$-decays.
\medskip

\noindent {\it Acknowledgements}:
Research partly supported by EU contract EURIDICE (HPRN-CT2002-00311),
PNL2005-41, FPA2002-00748 and the Ramon y Cajal Program.

\end{document}